\magnification 1200
Time-reversal asymmetry in Cont-Bouchaud stock market model
\bigskip
Iksoo Chang* and Dietrich Stauffer

\bigskip
Institute for Theoretical Physics, Cologne University, D-50923 K\"oln, Euroland

*Present and permanent address: Department of Physics, Pusan National 
University, Pusan 609-735, Korea; chang@random.phys.pusan.ac.kr

\bigskip
\noindent
The percolation model of stock market speculation allows an asymmetry (in the
return distribution) leading
to fast downward crashes and slow upward recovery. We see more small upturns
and more intermediate downturns.

\bigskip

Many microscopic models of speculation in  economic markets have been published
since Stigler's first Monte Carlo simulation of 1964 [1]. Most of them give
symmetric return distributions: The probability of gaining $r$ percent is 
the same as the probability of a $r$ percent loss, during the same time period.
This symmetry is also roughly found in reality. However, in the crashes of
1929 and 1987 on Wall Street (as well as in March and April 2001), the fast 
crash was followed by a slower recovery
[2], and a more sophisticated analysis of this lack of time-reveral invariance
was published in [3]. We now try to put such an effect into the percolation 
model of Cont and Bouchaud [4] which was modified in many ways to give specific
effects as observed in reality.

In the standard model, which we simulate here on triangular lattices, every
lattice site is randomly either empty (in our case with probability 1/2) or 
occupied by an agent (also with probability 1/2). The resulting percolation 
clusters act as rigidly coupled sets of investors: They either all sell together
with probability $a$, or buy with probability $a$, or are inactive with
probability $1-2a$, during one time interval. The amount of trade is 
proportional to the number of agents in the cluster, and the relative price 
change $r$ varies as the difference between supply and demand, summed over all
active clusters. This model gave [4] a symmetric distribution of returns $r$ 
with smooth maximum at $r=0$ and a power-law tail for large returns, except for 
finite-size limitations of $r$ and for a more Gaussian behaviour at large $a$. 
Modifications gave volatility clustering, outliers for very large $r$, and 
an asymmetry between sharp peaks and smooth valleys of price versus time.
This asymmetry, however, is different from the one between past and future [4], 
on which we concentrate on now. We average over all times, while Lillo and 
Mantegna found asymmetries when separating crash periods from rally periods [5].

Instead of giving the same probability $a$ to buy and sell, we assume that
after a downward movement of the price agents
have a higher probability to sell than
to buy (``panic''), while after an upward movement the probability to buy is
higher than that to sell. However, the panic effect is taken as ten times 
stronger than the influence of a price increase. In this case, prices show a
downward trend apart from fluctuations, which gives an asymmetry trivially but
is unrealistic. Therefore we also take into account a fundamentalist restoring
force [6] like in the Ornstein-Uhlenbeck process: High prices cause more
people to sell than to buy, while low prices increase the buying probability.
Now after $10^3$ time steps the prices settle down to a stationary plateau 
below the initial price, and we average only over the second half of the 
simulation.

The semilogarithmic plot of Fig.1 shows a maximum in the center, a power law
for intermediate $r$, and an exponential decay for large $r$. More interesting
are the same data in the center, plotted in Fig.2 on expanded scales. The 
maximum is now visibly shifted to the right: More price increases for small
$r$. But for negative $r$ a small bump is seen in the power-law regime: More
price decreases at some intermediate $r$. (Since the overall prices are roughly
constant, any increase of the probability for positive $r$ must be matched by 
an increase also
at some negative $r$.) It would be nice to see if high-precision data [7]
of real markets, looked at more precisely as in Fig.2, will give similar
effects as in our simulations based on $3.2 \times 10^9$ price changes.

We thank J.-F. Muzy, M. Ausloos, S. Solomon, D. Sornette and Y.-C. Zhang for 
helpful comments, DFG and KOSEF for supporting the visit of IC to Cologne,
the J\"ulich Supercomputer Centre for time on their Cray-T3E, and the High
Performance Supercomputing Centre of Pusan National University.

\bigskip
\parindent 0pt
[1]  H. Levy, M. Levy and S. Solomon, {\it Microscopic Simulation of Financial
Markets}, Academic Press, New York 2000.

[2] D. Davis, and C. Holt,  {\it
  Experimental Economics}, Princeton University Press, Princeton 1993 ;
J. Kagel, J. and A. Roth, eds., {\it  Handbook of Experimental Economics},
Princeton University Press, Princeton 1995.
S. Solomon, priv. comm.; M. Ausloos, priv. comm., Y.C. Zhang, Physica A
269 (1999) 30.

[3] A. Arneodo, J.F. Muzy and D. Sornette, Eur. Phys. J. B 2 (1998) 277; J.-F. 
Muzy, D. Sornette, J. Delour and A. Arneodo, Quantitative Finance 1 (2001) 131.

[4] R. Cont and J.P. Bouchaud, eprint cond-mat/9712318 = Macroeconomic Dynamics
 4 (2000) 170; D. Stauffer, Adv. Complex Syst. 4 (2001) No.1.

[5] F. Lillo, R.N. Mantegna, Eur. Phys. J. B 15 (2000) 603;
Phys. Rev. E 62 (2000) 6126.

[6] I. Chang and D. Stauffer, Physica A 264 (1999) 294.

[7] P. Gopikrishnan, V. Plerou, L.A.N. Amaral, M. Meyer, H.E. Stanley,
Phys. Rev. E 60 (1999) 5305.

\bigskip
Captions:

Fig.1: Histogram of returns, in arbitrary units, for activity $a = 0.05$
averaged over nearly a million $301 \times 301$ critical percolation
lattices, each followed over 5000 time steps, showing an overall symmetric 
behaviour. At the ends of the tails, consecutive bins were combined.

Fig.2: This expanded central region of Fig.1 shows clearly the desired
asymmetry between left (downturns) and right (upturns). The dashed line 
gives the standard model without panicky or fundamentalist agents. 
\end